\documentclass{article}
\usepackage{spconf}
\usepackage{cite}
\usepackage{amsmath,amssymb,amsfonts}
\usepackage{algorithmic}
\usepackage{graphicx}
\usepackage{textcomp}
\usepackage{xcolor}
\usepackage{booktabs}
\usepackage[hidelinks]{hyperref}

\def\BibTeX{{\rm B\kern-.05em{\sc i\kern-.025em b}\kern-.08em
    T\kern-.1667em\lower.7ex\hbox{E}\kern-.125emX}}

\title{DriftTTS: Few-Step Text-to-Speech Without Distillation via Distribution-Matching Drift}
\name{Mohammad Nur Hossain Khan$^{\star}$, Subrata Biswas$^{\dagger}$, Bashima Islam$^{\star}$}
\address{$^{\star}$University of Massachusetts Amherst, MA, USA \quad
         $^{\dagger}$Worcester Polytechnic Institute, MA, USA}

\begin{document}
\maketitle

\begin{abstract}
Few-step neural text-to-speech models often rely on shortened diffusion or flow-matching schedules, or on distillation from pretrained multi-step teachers. To avoid these dependencies, we present DriftTTS, a few-step mel-spectrogram generator trained without a generative teacher, distillation, or adversarial discrimination. DriftTTS uses a distribution-matching drift objective in a mel-domain feature space defined by raw mels and a frozen masked-autoencoder encoder pretrained on the same LJSpeech training split. On-policy rollout trains the decoder on its own intermediate states and supports inference up to the trained rollout depth. On LJSpeech, DriftTTS at NFE=4 achieves 3.87~dB MCD and 3.7\% WER, compared with 3.85~dB and 3.4\% for Matcha-TTS. In a fully paired blind listening test, DriftTTS obtains 4.18 MOS, compared with 3.96 for Matcha-TTS and 4.22 for ground truth. These results demonstrate competitive few-step synthesis without a pretrained generative teacher. Code can be found at \url{https://github.com/BASHLab/driftTTS.git}
\end{abstract}


\begin{keywords}
text-to-speech, drifting models, few-step generation, distillation-free
synthesis, flow matching
\end{keywords}

\section{Introduction}
\label{sec:intro}

Non-autoregressive neural text-to-speech (TTS) has moved from deterministic regression~\cite{ren2021fastspeech2} and end-to-end adversarial training~\cite{kim2021vits} toward iterative generative models. Grad-TTS~\cite{popov2021gradtts} integrates a diffusion SDE, and Matcha-TTS~\cite{mehta2024matcha} integrates a conditional flow-matching~\cite{lipman2023flowmatching} ODE while remaining competitive with only a few integration steps. These models reach high synthesis quality, but their inference cost grows with the number of function evaluations (NFE). Their training also shares a common structure. The network regresses a known per-sample target, such as the added noise or the velocity along a prescribed noise to data path, and the sampler integrates the learned field at inference.

A common route to one-step synthesis is distillation. CoMoSpeech~\cite{ye2023comospeech} distills a consistency model from a diffusion teacher, while ProDiff~\cite{huang2022prodiff} progressively distills an $N$-step diffusion model into fewer-step students. Teacher-free alternatives include FastSpeech~2~\cite{ren2021fastspeech2}, which regresses acoustic targets, VITS~\cite{kim2021vits}, which uses variational and adversarial training, and FlashSpeech~\cite{ye2024flashspeech}, which trains a one- or two-step latent consistency model from scratch with adversarial consistency training. Consistency models can also be trained without a teacher~\cite{song2023consistency}. These approaches still rely on per-sample regression, adversarial discrimination, or consistency along a prescribed trajectory. Teachers require multi-stage training, discriminators introduce an additional network and min-max objective, and consistency constraints tie generation to a fixed noise$\to$data path. Whether a competitive few-step acoustic model can be trained with none of these ingredients has, to our knowledge, not been examined for text-to-speech.


We propose DriftTTS, a mel-spectrogram acoustic model built on Drifting Models~\cite{deng2026drifting}. This recent generative framework replaces per-sample regression targets with a \emph{distribution-level} matching force, termed the drift. Given batches of generated and real samples embedded in a feature space, the drift attracts generated samples toward the data and repels them from one another, and the generator is updated to follow it. No teacher, discriminator, or prescribed trajectory is involved. A related formulation has recently been applied to speech \emph{enhancement}~\cite{xu2026speech}. Applying the drift objective to TTS requires several adaptations. Because each text is paired with only one recording, we construct a local condition-specific positive pool around the target segment. We further extend the original one-step formulation to few-step generation and define the drift in a mel-domain feature space that captures perceptually relevant differences.


DriftTTS adapts the drifting framework to TTS by optimizing the decoder with the drift objective. For each conditioning segment, the paired target and its noise-augmented views form a local positive pool, while generated samples from other utterances in the batch provide negatives. To move beyond a single step, we train the decoder by on-policy rollout. The model is unrolled for a random number of its own steps up to a fixed training depth, and the drift objective is applied at the visited state, so no noise-to-data trajectory is prescribed. At inference, the same trained model can be run at any number of evaluations up to that depth, which makes the step count an explicit control rather than a property fixed at training time. The drift is computed in the feature space of a frozen masked autoencoder that we pretrain on the same LJSpeech training split, together with the raw mel representation. This network is used only to construct the objective and is discarded at inference. No generative teacher, distillation stage, or discriminator is used at any point. To our knowledge, DriftTTS is the first application of the drift objective to TTS. Our contributions are:


\vspace{-0.5em}
\begin{enumerate}
\itemsep0em
\item \textbf{DriftTTS}, a TTS model whose decoder is trained entirely with a distribution-matching drift objective in a mel-domain perceptual feature space, without a generative teacher, distillation, or adversarial training.
\item A \textbf{few-step on-policy training scheme }that extends the one-step drift generator to iterative refinement with explicit NFE control at inference, where step count is the largest quality lever in our experiments.
\item A \textbf{vocoder-controlled evaluation} against Matcha-TTS on LJSpeech, centred on a fully paired blind listening test in which every listener rated every system on the same utterances and results are analysed both by subject and by utterance. Under this protocol DriftTTS at NFE=4 reaches naturalness comparable to Matcha-TTS and to ground truth (4.18 against 3.96 and 4.22 MOS), and sampling the same trained model with four evaluations instead of one raises MOS by 1.18 points, significant under both analyses.
\end{enumerate}
\vspace{-1em}
\section{DriftTTS}
\label{sec:method}

\begin{figure*}
    \centering
    \includegraphics[width=0.95\linewidth]{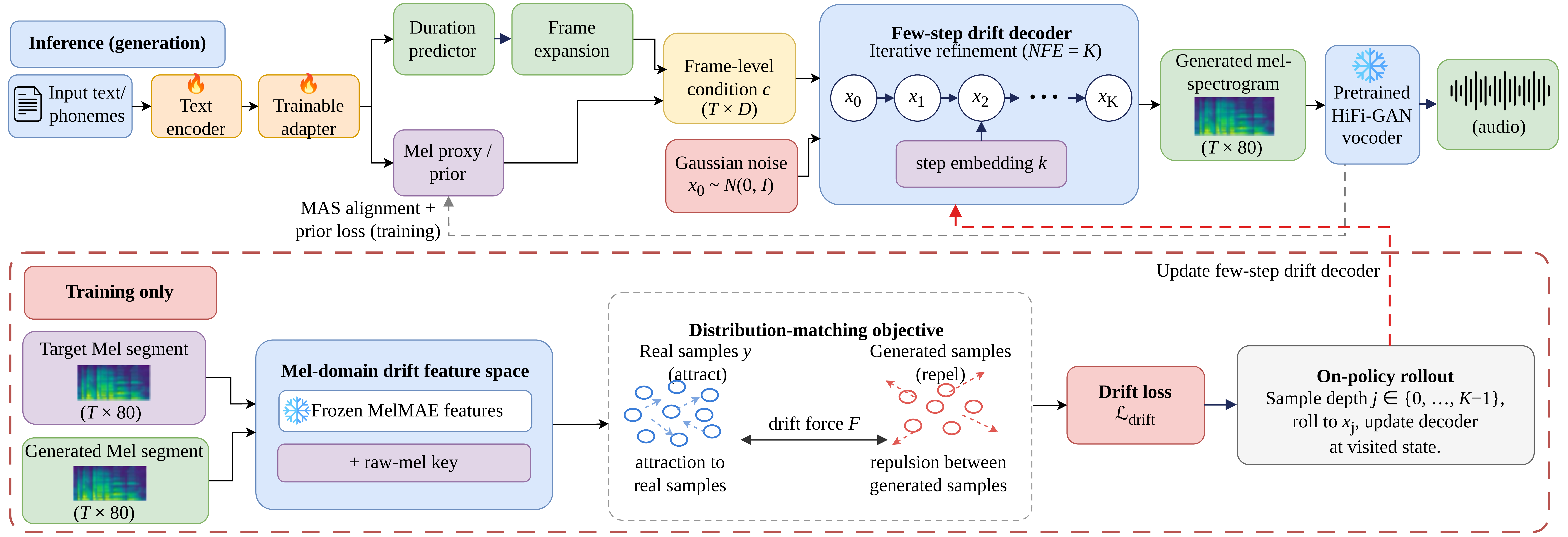}
    \caption{Overview of DriftTTS. Text-conditioned Gaussian noise is iteratively refined by a few-step drift decoder to generate mel-spectrograms, which are converted to waveform by HiFi-GAN. During training, the decoder is optimized with an on-policy distribution-matching drift objective computed in a mel-domain feature space using real and generated samples.}
    \label{fig:overview}
    \vspace{-1em}
\end{figure*}

DriftTTS replaces the prescribed noise-to-data field used by diffusion and flow-matching TTS with the distribution-matching objective of Drifting Models~\cite{deng2026drifting}. Let $g_i$ denote generated features and $p_j$ real features. We form generated-to-real distances $d^{+}_{ij}=\|g_i-p_j\|_2$ and generated-to-generated distances $d^{-}_{ik}=\|g_i-g_k\|_2$, $k\neq i$. Each distance matrix is normalized by its own batch mean, and at each scale $R\in\{0.2,0.05,0.02\}$ we construct cross and self affinities using
\begin{equation}
a_{ij}^{(R)}
=
\sqrt{
\operatorname{softmax}_{\rm row}\!\left(-\frac{\tilde d_{ij}}{R}\right)
\operatorname{softmax}_{\rm col}\!\left(-\frac{\tilde d_{ij}}{R}\right)
}.
\label{eq:affinity}
\end{equation}
The same construction is applied to both distance matrices, with the diagonal masked for the generated-to-generated self-affinity. The resulting raw force is
\begin{equation}
\widehat F_i^{(R)}
=
\sum_j a_{ij}^{+,(R)}p_j
-
\sum_{k\neq i}a_{ik}^{-,(R)}g_k,
\label{eq:force}
\end{equation}
where the two terms are the attractive and repulsive components, respectively. Following~\cite{deng2026drifting}, each scale is normalized by its batch RMS,
\vspace{-1em}

\begin{equation}
\resizebox{.90\columnwidth}{!}{$\displaystyle
\lambda_R=
\left(
\frac{1}{BD_R}
\sum_{i=1}^{B}
\|\widehat{F}_i^{(R)}\|_2^2
\right)^{1/2},
\qquad
F_i=\sum_R\frac{\widehat{F}_i^{(R)}}{\lambda_R+\epsilon}, $}
\label{eq:force_norm}
\end{equation}
where $D_R$ is the feature dimensionality. Let $E(\cdot)$ denote the frozen mel-domain feature mapping and $f_{\theta,\phi}(z_i,x_i)$ the generated mel-spectrogram, with decoder parameters $\theta$ and text-side parameters $\phi$. Training uses stop-gradient regression toward the displaced feature representation,
\begin{equation}
\resizebox{.90\columnwidth}{!}{$\displaystyle
\mathcal{L}_{\mathrm{drift}} = \frac{1}{B}\sum_{i=1}^{B}\|E(f_{\theta,\phi}(z_i,x_i)) - \mathrm{sg}[E(f_{\theta,\phi}(z_i,x_i)) + F_i]\|_2^2
$}
\end{equation}

The force $F_i$ is defined in feature space and is pulled back through the frozen encoder $E$ to optimize the text encoder and decoder, while $E$ itself remains fixed. Because the displacement is normalized, the numerical value of $\mathcal{L}_{\rm drift}$ is not used as a convergence measure. We instead monitor the pre-normalization batch RMS $\lambda_R$, which decreases as the raw drift approaches equilibrium.
\vspace{-0.7em}
\subsection{Training}
\label{sec:training}

The text encoder, duration predictor, and monotonic alignment follow the Grad-TTS/Matcha-TTS framework. Given target mel-spectrogram $y$, the text encoder produces token representations and a mel-domain proxy $\mu$; Monotonic Alignment Search \cite{kim2020glowtts} determines the text-to-frame alignment, and the text side is trained with $\mathcal{L}_{\rm prior}=\|\mu_{\rm frame}-y\|_2^2$, together with the duration-prediction loss.

The acoustic generator is trained end-to-end with the drift objective and no reconstruction loss on the generated mel. Training uses random 192-frame segments. For each conditioning segment, we generate $g=8$ samples sharing the same text condition. The positive pool contains the paired target segment and seven Gaussian-perturbed views ($\sigma=0.05$ in normalized mel space), providing a local conditional neighborhood for distribution matching.

The drift uses raw 80-bin mels together with a frozen MelMAE pretrained on the LJSpeech training split. MelMAE uses $16\times16$ patches, hidden dimension 256, 60\% masking, a six-layer encoder, and a two-layer reconstruction decoder~\cite{he2022mae}. It is pretrained for 30k steps with AdamW ($3\times10^{-4}$ learning rate, weight decay 0.05, cosine decay). The 4.8M-parameter encoder is then frozen, with block-3 and block-6 activations used as drift features, while the reconstruction decoder is discarded. MelMAE is used only during training and adds no inference cost.

To train a few-step generator, we use on-policy rollout. For maximum training depth $K$, we sample $j\sim\mathcal{U}\{0,\ldots,K-1\}$, roll the current model forward for $j$ steps without gradient computation, $x_{i+1}=D_\theta(x_i,c,i), \qquad i<j,$ and evaluate the drift objective at the visited state $x_j$. Thus, the model is trained on states induced by its own current rollout rather than on an externally prescribed noise-to-data trajectory or teacher-derived intermediate targets. Empirically, the repulsive component is concentrated at early rollout states. Once generated samples contract, its gradient becomes nearly identical to attraction-only. Repulsion acts primarily as an early-rollout shaping term. 
\vspace{-0.5em}
\subsection{Inference and architecture}
\label{sec:inference}
\vspace{-0.5em}
At inference, the text encoder and duration predictor produce the frame-level conditioning sequence $c$. Generation starts from
$x_0 \sim \mathcal{N}(0,I)$, and the decoder performs replacement updates
$x_{k+1}=D_\theta(x_k,c,k), \qquad k=0,\ldots,K-1.$ Here $D_\theta$ directly predicts the next mel state rather than a residual, and a learned step embedding identifies $k$. The same model can be evaluated at any NFE up to its trained depth $K$; our primary configuration uses $K=4$. The decoder is a six-layer convolutional Transformer operating on 80-bin mel states, with hidden dimension 256, four attention heads, and cross-attention to 192-dimensional frame-level text representations. A two-layer Transformer adapter is applied to the text representation, and the text encoder, adapter, duration predictor, mel projection, and decoder are trained jointly. The decoder contains 7.37M parameters and the full acoustic model contains 15.70M trainable parameters. The generated mel-spectrogram is converted to waveform using a pretrained HiFi-GAN vocoder \cite{kong2020hifigan}.
\vspace{-1em}
\section{Experiments}

\subsection{Experimental Setup}

We evaluate DriftTTS on LJSpeech using the official Matcha-TTS partition of 12,500/100/500 train/validation/test utterances. Audio is represented using 80-bin mel-spectrograms. DriftTTS is trained for 60k updates on a single NVIDIA L40S using AdamW with learning rate $10^{-4}$ and cosine decay to 5\% of the initial rate. Training uses 192-frame mel segments with $g=8$ generated samples and $n=8$ positive samples per conditioning segment.

We compare against Matcha-TTS~\cite{mehta2024matcha}, a 18.2M parameter model, using its official pretrained LJSpeech checkpoint and evaluate NFE $\in\{1,2,4\}$. DriftTTS uses $K=4$ as the primary configuration and is evaluated at NFE $\in\{1,2,4\}$ in the main comparison. We also evaluate the official Grad-TTS~\cite{popov2021gradtts} checkpoint at NFE=10; lower-NFE settings produced substantially degraded synthesis quality and were not competitive in our evaluation. All synthesized systems use the same pretrained HiFi-GAN vocoder. We report mel-cepstral distortion (MCD-DTW) \cite{kubichek1993mcd} on resynthesized waveforms, word error rate (WER) using Whisper-medium \cite{radford2023whisper}, UTMOS \cite{saeki2022utmos}, and a fully paired blind mean-opinion-score (MOS) test.
\vspace{-0.5em}
\subsection{Main Results}
\vspace{-0.5em}
Table~\ref{tab:main} summarizes the objective and subjective results. At NFE=4, DriftTTS reaches 3.87~dB MCD and 3.7\% WER, compared with 3.85~dB and 3.4\% for Matcha-TTS. DriftTTS therefore approaches the spectral fidelity and intelligibility of the flow-matching baseline while using fewer parameters.

Matcha-TTS at NFE=4 obtains UTMOS 4.36, exceeding the 4.20 score of ground-truth mels resynthesized through the same vocoder, while Matcha-TTS at NFE=1 matches DriftTTS-4 in UTMOS (4.25) despite substantially higher MCD (4.79 vs.\ 3.87~dB). WER is similarly non-discriminative for quality at this operating point. DriftTTS changes only from 3.6\% WER at NFE=1 to 3.7\% at NFE=4, while MOS rises from 3.00 to 4.18. Thus, WER captures intelligibility but not the large naturalness difference between these conditions. These differences suggest that UTMOS remains informative but may not fully resolve fine-grained naturalness differences in this high-quality regime, motivating complementary human evaluation. To evaluate perceived naturalness directly, we conduct a fully paired blind listening test with 15 raters, 20 randomly selected utterances, and 5 systems. All raters are recruited through institutional advertisement with no compensation. Every listener rates all systems on the same utterances using a five-point naturalness scale. This produces 300 ratings per system and 1500 ratings in total. DriftTTS at NFE${=}4$ obtained a MOS of $4.18\pm0.16$, compared with $3.96$ for Matcha-TTS and $4.22$ for ground-truth recordings. Both DriftTTS and Matcha-TTS exceed Grad-TTS-10 at 3.55 while using fewer evaluations.

\begin{table}[t]
\caption{Evaluation results on LJSpeech. All synthesized systems use the same HiFi-GAN vocoder, and WER is computed using Whisper-medium. Listening-test statistics are reported in Table~\ref{tab:mos}.}
\label{tab:main}
\centering
\small
\resizebox{\linewidth}{!}{
\begin{tabular}{lccccc}
\toprule
System & NFE & WER$\downarrow$ & MCD$\downarrow$
       & UTMOS$\uparrow$ & MOS$\uparrow$ \\
\midrule
Ground truth        & -- & 0.029 & 0.00 & 4.37 & $4.22$ \\
GT mel $\rightarrow$ vocoder
                    & -- & -- & 2.14 & 4.20 & -- \\
\midrule
Matcha-TTS          & 1  & 0.042  & 4.79 & 4.25 & -- \\
Matcha-TTS          & 4  & 0.034 & 3.85 & 4.36 & $3.96$ \\
\midrule
Grad-TTS            & 10 & 0.043 & 4.17 & 3.76 & $3.55$ \\
\midrule
DriftTTS            & 1  & 0.036 & 4.89 & 3.16 & $3.00$ \\
DriftTTS            & 2  & 0.041    & 4.08 & 4.01 & -- \\
\textbf{DriftTTS}   & \textbf{4}
                    & \textbf{0.037}
                    & \textbf{3.87}
                    & \textbf{4.25}
 & $\mathbf{4.18}$ \\
           
\bottomrule
\end{tabular}}
\vspace{-1em}
\end{table}



The effect of the number of drift steps, in contrast, is clear. Increasing DriftTTS from NFE${=}1$ to NFE${=}4$ raises MOS from 3.00 to 4.18, a gain of 1.18 points. The difference is significant both by subject ($p=0.001$) and by item ($p<0.001$). The improvement is therefore consistent across both listeners and utterances. We additionally measure component-level inference latency for 20 utterances with an average length of 487 and repeated 10 times on an NVIDIA RTX PRO 6000 with batch size one. At matched NFE=4, the DriftTTS decoder requires 7.08~ms compared with 11.22~ms for Matcha-TTS, corresponding to a $1.58\times$ decoder speedup. Excluding the vocoder, the complete acoustic path requires 10.35~ms and 14.53~ms, respectively.

\begin{table}[t]
\caption{Blind naturalness MOS and paired comparisons. Two-sided paired t-tests are performed both across subjects and across utterances. CI was measured over subjects' mean rating.}
\label{tab:mos}
\centering
\small
\begin{tabular}{lcc}
\toprule
System & NFE & MOS $\pm$ 95\% CI \\
\midrule
Ground truth            & -- & $4.22 \pm 0.17$ \\
\textbf{DriftTTS}       & 4  & $\mathbf{4.18 \pm 0.16}$ \\
Matcha-TTS              & 4  & $3.96 \pm 0.18$ \\
Grad-TTS                 & 10 & $3.55 \pm 0.15$ \\
DriftTTS                & 1  & $3.00 \pm 0.21$ \\

\midrule
\multicolumn{3}{l}{\emph{Paired comparisons}}\\
\midrule
\multicolumn{2}{l}{DriftTTS-4 vs.\ DriftTTS-1}
& $p=.001$ / $p<.001$ \\
\multicolumn{2}{l}{DriftTTS-4 vs.\ Matcha-TTS-4}
& $p=.31$ / $p=.044$ \\
\multicolumn{2}{l}{DriftTTS-4 vs.\ ground truth}
& $p=.72$ / $p=.79$ \\
\bottomrule
\end{tabular}
\vspace{-1em}
\end{table}

\begin{table}[t]
\centering
\caption{Ablation of the drift objective and feature space. All models use the same architecture, $K=4$ rollout, and 60k training updates. Lower is better for all metrics.}
\label{tab:mel_ablation}
\small
\resizebox{\linewidth}{!}{
\begin{tabular}{lccc}
\toprule
Configuration & MCD $\downarrow$ & MCD Std. $\downarrow$ &
Worst MCD $\downarrow$ \\
\midrule
MelMAE + raw mel, drift & \textbf{3.873} & \textbf{0.457} &
\textbf{4.78} \\
Raw mel only, drift     & 4.065 & 0.575 & 5.84 \\
MelMAE + raw mel, L2    & 4.309 & 0.616 & 6.91 \\
\bottomrule
\end{tabular}}
\vspace{-1em}
\end{table}

\begin{table}[t]
\centering
\caption{Effect of training rollout depth on synthesis quality and learned dynamics. Evaluated at $NFE=K$.}
\label{tab:training_depth}
\small
\begin{tabular}{c c c c c}
\toprule
Trained $K$ &
MCD $\downarrow$ &
WER $\downarrow$ &
UTMOS $\uparrow$ & \\
\midrule
2 & 4.76 & 0.0705 & 2.818  \\
4 & \textbf{3.87} & \textbf{0.0373} & \textbf{4.253} \\
\bottomrule
\end{tabular}
\vspace{-1em}
\end{table}
\vspace{-0.5em}
\subsection{Ablations and Analysis}

\noindent\textbf{Drift objective and feature space.}
Table~\ref{tab:mel_ablation} isolates the contributions of the training objective and feature representation. Removing the MelMAE features and using raw mels alone increases MCD from 3.873 to 4.065~dB and increases worst-utterance MCD from 4.78 to 5.84~dB. Replacing the drift objective with direct $\ell_2$ regression further increases MCD to 4.309~dB and yields the largest variance and worst-case distortion. Both the distribution-matching objective and the combined learned/raw-mel feature space therefore contribute to synthesis quality. The raw-mel-only drift model also outperforms the MelMAE-based $\ell_2$ model (4.065 vs.\ 4.309~dB), showing that the distribution-matching objective improves quality even without pretrained features.

\noindent\textbf{Training rollout depth}
Table~\ref{tab:training_depth} shows that training depth strongly affects synthesis quality. Increasing $K$ from 2 to 4 reduces MCD from 4.76 to 3.87~dB and WER from 7.05\% to 3.7\%, while increasing UTMOS from 2.818 to 4.253. These results motivate $K=4$ as our primary training configuration.

\noindent\textbf{Generalization beyond the training window.}
The 500 test utterances span 100--867 frames ($0.52\times$--$4.52\times$ the training window), with 474/500 exceeding it. MCD improves with length, with a slope of $-0.0458$~dB per 100 frames (95\% CI $[-0.075,-0.017]$); the longest quartile outperforms the shortest by 0.236~dB ($p=0.005$). Thus, the fixed training window does not cause degradation on longer utterances.


\noindent\textbf{Behavior beyond the trained rollout depth}
The on-policy objective samples $j\sim\mathcal{U}\{0,\ldots,K-1\}$, so a model trained with depth $K$ only optimizes step indices $0,\ldots,K-1$. For the $K=2$ model, MCD increases from 4.76~dB at NFE=2 to 5.81~dB at NFE=4 when inference extends beyond the trained depth. In contrast, a model trained with $K=4$ remains usable through four evaluations, with the best MCD being 3.87~dB. We therefore restrict inference to $\mathrm{NFE}\leq K$.
\vspace{-1em}

\section{Conclusion}
\vspace{-0.5em}
We introduced DriftTTS, a few-step TTS acoustic model trained with a distribution-matching drift objective without a pretrained generative teacher, distillation, or adversarial training. On LJSpeech, DriftTTS achieves objective quality close to Matcha-TTS and a MOS of 4.18 at NFE=4, while ablations confirm the importance of both the drift objective and the combined mel-domain feature representation. These results show that competitive few-step speech synthesis can be achieved without a pretrained generative teacher or distillation.

\vfill\pagebreak

\bibliographystyle{IEEEtran}
\bibliography{drifttts}

\end{document}